\documentclass[journal]{IEEEtran}

\ifCLASSINFOpdf
\else
   \usepackage[dvips]{graphicx}
\fi
\usepackage{url}

\usepackage[utf8]{inputenc} 
\usepackage[T1]{fontenc}
\usepackage{url}              
\usepackage{cite}             

\usepackage{graphicx}
 \usepackage{amsmath}
\usepackage{amssymb}
\usepackage{bm}
\def\t#1{\bm{\tilde{#1}}}
\def\ti#1{{\tilde{#1}}}
\def\bb#1{{\mathbb #1}}

\begin{document}

\title{Deep Unfolded Simulated Bifurcation for Massive MIMO Signal Detection}

\author{Satoshi Takabe, \IEEEmembership{Member, IEEE}
\thanks{The author thanks Mr. Ryo Hagiwara for commenting on the manuscript.
The codes are available at \protect\url{https://github.com/s-takabe/unfolded_simbif}. 
This work was partly supported by  JSPS KAKENHI Grant Numbers 22H00514, 22K17964.}
\thanks{S. Takabe is with Tokyo Institute of Technology, Ookayama, Tokyo 152-8550 Japan (e-mail: takabe@c.titech.ac.jp).}}


\maketitle

\begin{abstract}
Multiple-input multiple-output (MIMO) is a key ingredient of next-generation wireless communications. Recently, various MIMO signal detectors based on deep learning techniques or quantum(-inspired) algorithms have been proposed to improve the detection performance compared with conventional detectors. This paper focuses on the simulated bifurcation (SB) algorithm, a quantum-inspired algorithm.  This paper proposes two techniques to improve its detection performance. The first is modifying the algorithm inspired by the Levenberg–Marquardt algorithm
 to eliminate local minima of the maximum likelihood detection. The second is the use of deep unfolding, a deep learning technique to train the internal parameters of an iterative algorithm. We propose a deep-unfolded SB by making the update rule of SB differentiable. The numerical results show that these proposed detectors significantly improve the signal detection performance in massive MIMO systems.
\end{abstract}


\IEEEpeerreviewmaketitle

\section{Introduction}

\IEEEPARstart{M}{ultiple}-input multiple-output (MIMO) is a key ingredient of next-generation wireless communications~\cite{MUMIMO,Yang}. In massive MIMO systems, the exact maximum likelihood (ML) detection is computationally intractable, and the performance of conventional detectors such as a minimum mean-squared error (MMSE) detector~\cite{MMSE} degrades seriously. 
To overcome the difficulty, a number of approximate ML detectors have been developed. 
In particular, recent MIMO signal detectors are divided into two classes: deep learning (DL)-based detectors and quantum(-inspired) detectors.   

DL techniques have been applied to various fields of signal processing.  
In particular, deep unfolding (DU)~\cite{LISTA,DU} is a powerful tool for constructing a trainable signal detector for signal processing~\cite{Bala,DU2}. 
We embed some trainable internal parameters by combining DU with an existing iterative algorithm.
We then train these parameters by conventional DL techniques such as back-propagation. 
DU is thus model-based learning and useful to utilize the mathematical knowledge of the research field. 
DU has been applied to various signal processing tasks such as compressed sensing~\cite{TISTA} and signal detection in wireless communications~\cite{DetNet,TPG2,He2}.

Along with DL techniques, the use of quantum(-inspired) computation has attracted attention in optimization and signal processing~\cite{IM}. 
Quantum computation techniques such as quantum annealing~\cite{Ki,QA1,QA2} and coherent Ising machine~\cite{CIM} have been applied to MIMO signal detection and show reasonable detection performance. 
In addition, quantum-inspired computation has been attractive because it is fast classical computation without  the restriction of computational  resources.       
For example, simulated bifurcation (SB)~\cite{SB1} is a classical dynamical system solving a quadratic unconstrained  binary optimization problem (QUBO) inspired by quantum nonlinear parametric oscillators. 
It was reported that the SB-based MIMO detector suffers from an error floor, although SB is helpful for solving  other huge combinatorial optimizations~\cite{SB3}. 
The error floor is sometimes observed in other quantum(-inspired) MIMO detectors~\cite{CIM}.

The aim of this letter is twofold. 
First, we attempt to improve the error floor of the SB-based MIMO detector by modifying it. We will numerically show that the modification eliminates local minima that cause the detection error and significantly improves the detection performance. 
Secondly, we combine SB with DU to further improve the detection performance with fewer  iterations. 
We introduce a novel deep unfolded SB detector as a differentiable iterative algorithm and train some internal parameters that control its performance. Some numerical experiments are conducted to show the performance improvement by the training process based on DU.

This letter is organized as follows. 
Section~\ref{sec_set} describes the MIMO system model.
In Sec.~\ref{sec_SB}, we describe an existing SB-based detector. 
We propose a technique to eliminate local minima in SB-based detection in Sec.~\ref{sec_eli}.
Section~\ref{sec_DUSB} describes the deep unfolded SB detector and shows its detection performance.
Section~\ref{sec_con} summarizes this letter.

\section{Model Setting}\label{sec_set}
In this section, we describe the channel model.
The number of transmitting and receiving antennas is denoted by $N_t$ and $N_r$, respectively. 
For simplicity, it is supposed that no precoding is used and that the channel state information is known perfectly.

Let $\t{x} := [\ti{x}_1,\ti{x}_2,\dots,\ti{x}_{N_t}]^{T} \in \t{{\mathbb S}} ^{N_t}$ be 
a vector whose element $\ti{x}_j$ ($j=1,\dots,{N_t}$) is a transmit symbol  from the $j$-th antenna.
The set $\t{\bb{S}} \subset \bb{C}$ represents a signal constellation. 
Assuming a flat Rayleigh fading channel,  the received signal  $\t{y} := [\ti{y}_1,\ti{y}_2,\dots,\ti{y}_{N_r}]^{T} \in \bb{C}^{N_r}$ 
is given by 
$
  \t{y} = \t{H}\t{x} + \t{w}, 
$
where $\t{H} \in \bb{C}^{{N_r} \times {N_t}}$ is a channel matrix and 
$\t{w}\in \bb{C}^{N_r}$ is a complex additive white Gaussian noise vector   
with zero mean and covariance matrix $\sigma_{w}^2\bm I$.

This channel model is equivalent to a real-valued channel defined by
\begin{equation}
\bm y = \bm H\bm x + \bm w, \label{eq_rec}
\end{equation}
where 
   \begin{align} \nonumber
      \bm y &:= \begin{bmatrix}
        \Re(\t{y}) \\
        \Im(\t{y})
        \end{bmatrix} \in \bb{R}^{M},\ 
      \bm H := \begin{bmatrix} \label{eq:H_real}
        \Re(\t{H}) & - \Im(\t{H})\\
        \Im(\t{H}) &  \Re(\t{H})\\
      \end{bmatrix}, \\ \nonumber
      \bm{x} &:= \begin{bmatrix}
        \Re(\t{{x}}) \\
        \Im(\t{{x}})
        \end{bmatrix} \in \bb{S}^{N}, \ 
      \bm{w} := \begin{bmatrix}
        \Re(\t{{w}}) \\
        \Im(\t{{w}})
        \end{bmatrix}\in \bb{R}^{M},
    \end{align}
 and $(N,M) := (2N_t,2N_r)$.
 In the following, we consider the QPSK modulation, i.e., $\mathbb{S}=\{1,-1\}$, in (\ref{eq_rec}).

\section{Simulated Bifurcation}\label{sec_SB}
Here, we briefly describe SB and its application to MIMO signal detection. 

\subsection{Basics of SB}
SB is a classical dynamical system inspired by quantum Kerr-nonlinear parametric oscillators~\cite{KPO}. 
SB approximately solves a minimization problem of the energy $E(\bm \sigma)$ defined by 
\begin{equation}
E(\bm \sigma) = \sum_{i,j=1}^{n} J_{ij}\sigma_i\sigma_j + \sum_{i=1}^n h_i\sigma_{i}, \label{eq_ene}
\end{equation}   
where $J_{ii}=0$, $J_{ij}=J_{ji}$ $(i,j=1,\dots,n)$ and $\bm \sigma=[\sigma_1,\dots,\sigma_n]^T\in\{1,-1\}^n$. 
The minimization problem of $E(\bm \sigma)$ is called QUBO and an NP-hard problem in general. 
In this letter, we focus on a variant of SB called ballistic SB~\cite{SB2} defined by 
\begin{align}
&\bm{\dot{x}}(t) = a_0\bm{y}(t), \label{eq_SB10}\\
&\bm{\dot{y}}(t) = -(a_0-a(t))\bm{x}(t) + c_0\left(\bm{J}\bm{x}(t)+\bm{h}\right), \label{eq_SB20}\\
&\mbox{if } |x_i(t)|>1, \mbox{then } x_i(t) =\mathrm{sign}(x_i(t)), y_i(t)=0, \nonumber
\end{align}   
where $\bm J = (J_{ij})$ and $\bm h = (h_i)$.  
In addition, $a_0,c_0$ are positive constant parameters, and $a(t)$ is a monotonically increasing  function of time $t\ge 0$. 
In the system, $x_i(t)$ is a continuous variable corresponding to $\sigma_i$ and $y_i(t)$ is an auxiliary variable. 
Starting from a random initial point $(\bm x(0),\bm y(0))$, $x_i(t)$ oscillates quickly when $a(t)\ll 1$ and approaches to either $1$ or $-1$ as $a(t)$ increases. 

Practically, 
the system (\ref{eq_SB10}) and (\ref{eq_SB20}) is solved numerically. 
Using the Euler method, we have an update rule given by
\begin{align}
&\bm{x}(k+1) = \bm{x}(k) + \Delta \bm y(k), \label{eq_SB1}\\
&\bm{y}(k+1) = \bm{y}(k)- \Delta\left[(1-a(k\Delta))\bm x(k) + c_0 \left(\bm J\bm x(k)+\bm h\right)\right], \label{eq_SB2}\\
&\mbox{if } |x_i(k)|>1, \mbox{then } x_i(k) =\mathrm{sign}(x_i(k)), y_i(k)=0, \nonumber
\end{align}   
where $\Delta$ is a time step. We set $a_0=1$ without loss of generality. 
The index $k=0,\dots, T-1$ represents the iteration step. 
We simply call this discretized version SB hereinafter. 
It is shown that, for a proper choice of $c_0$, the convergent $\bm x^\ast$ is a local minimum of the energy $E(\bm x)$. In other words, the convergent $\bm x^\ast$ satisfies 
\begin{equation}
\Delta E(x_i^\ast) =2x_i^\ast\sum_{j=1}^n J_{ij}x_j^\ast + h_ix_i^\ast \ge 0, \label{eq_loc}
\end{equation}   
for any $i=1,\dots,n$. 
Note that a convergent point of SB depends on an initial point and choice of parameters. In this sense, SB is an approximate algorithm for minimizing the energy $E(\bm\sigma)$.

\subsection{SB as MIMO signal detector}

Returning to the MIMO signal detection, we employ the ML estimator given by 
\begin{align}
\bm{\hat{x}}_{ML} &:= \arg\min_{\bm x\in\mathbb S^{N}} \frac{1}{2}\|\bm y-\bm H\bm x\|_2^2 \nonumber\\
&= \arg\min_{\bm x\in\mathbb S^{N}} \bm x^T\bm H^T\bm H\bm x - 2 \bm y^T \bm H \bm x , \label{eq_ML}
\end{align}   
indicating $\bm J = \bm H^T\bm H - \mathrm{diag}(\bm H^T\bm H)$ and $\bm h = -2\bm H^T\bm y$ in (\ref{eq_ene}). Then, SB is directly applicable to the MIMO signal detection~\cite{SB3}.
However, it was found that this SB detector called ML-SB exhibits an error floor in the high signal-to-noise (SNR)  region because SB stacks to local minima satisfying (\ref{eq_loc}). 

In~\cite{SB3}, the authors proposed the LMMSE-guided SB (G-SB) detector, which minimizes 
\begin{align}
\bm{\hat{x}}_{g} &:= \arg\min_{\bm x\in\mathbb S^{N}} \frac{1}{2}\|\bm y-\bm H\bm x\|_2^2 + \frac{\lambda_g}{2} \|\bm x -\bm x_{LMMSE}\|_2^2, \label{eq_g}
\end{align}   
where $\bm x_{LMMSE} = \bm H^T (\bm H\bm H^T+\sigma_w^2\bm I)^{-1} \bm y$ is the LMMSE solution and $\lambda_g\ge 0$ is a penalty coefficient. 
Although this modification increases the computational cost due to matrix inversion, it lowers the error floor~\cite{SB3}.

\section{Elimination of Local Minima}\label{sec_eli}

In this section, we propose another minimization strategy using an LMMSE-like matrix. 

The idea is based on some observations in trainable MIMO detectors. 
For $\bm x\in \mathbb{R}^N$, the gradient of $f_{ML}(\bm x) = \|\bm y-\bm H \bm x\|^2/2$ is given by 
$\nabla f_{ML} = -\bm H^T(\bm y-\bm H \bm x)$. 
It was reported that some gradient descent-based detectors, such as the TPG detector, show better detection performance by using an LMMSE-like matrix $\nabla f_{LM} = -\bm H^T (\bm H\bm H^T+\lambda \bm I)^{-1} (\bm y-\bm H \bm x)$ ($\lambda >0$) instead of $\nabla f_{ML}$~\cite{TPG2,He2}. 
The update rule using $\nabla f_{LM}$ is known as the Levenberg--Marquardt (LM) algorithm for nonlinear optimization~\cite{LM1,LM2}.

Here, we find the corresponding $\bm J$ and $\bm h$ to apply the LM algorithm to SB. 
Since we have  
$f_{LM}(\bm x) = \bm x^T \bm H^T (\bm H\bm H^T+\lambda \bm I)^{-1}\bm H \bm x/2 -  (\bm H^T (\bm H\bm H^T+\lambda \bm I)^{-1}\bm y)^T \bm x$, 
neglecting a constant term, we find 
\begin{equation}
\bm J = \bm U_\lambda \bm H- \mathrm{diag}(\bm U_\lambda\bm H), \bm h =  -2\bm U_\lambda\bm y, 
\label{eq_JH}
\end{equation}
where $\bm U_\lambda =\bm H^T (\bm H\bm H^T+\lambda \bm I)^{-1}$ is an LMMSE-like matrix with parameter $\lambda>0$.

\begin{table}
\caption{Values of objective functions of a toy example. We set $\lambda_g = 1$ for $f_G$ and $\lambda=1$ for $f_{LM}$.}
\label{table}
\small
\setlength{\tabcolsep}{3pt}
\centering
  \begin{tabular}{cccc}
    \hline
    $\bm x$  & $f_{ML}$  & $f_G$ &$f_{LM}$  \\
    \hline \hline
    $[1,1,1]^T$  &  $0.65$   & $3.91$ & $-61.8$\\
    $[-1,1,1]^T$  &  $4.53$  & $11.9$ & $-8.90$ \\
    $[1,-1,1]^T$  &  $5.55$  & $14.2$ & $-27.7$ \\
    $[1,1,-1]^T$  &  $4.01$  & $10.6$  & $-22.2$\\
    $[-1,-1,1]^T$  &  $5.10$  & $13.6$ & $24.2$ \\
    $[1,-1,-1]^T$  &  $4.95$ & $13.0$ & $11.0$  \\
    $[-1,1,-1]^T$  &  $4.20$  & $11.4$ &  $29.5$ \\
    $[-1,-1,-1]^T$  &  $0.91$  & $5.23$  & $61.8$\\ 
    \hline
  \end{tabular}\label{tab1}
\end{table}

To see the effectiveness of the use of $\nabla f_{LM}$, we here show a simple toy example.
As a real-valued noisy MIMO channel with $N=M=3$, we set  
\begin{equation*}
\bm H = \begin{bmatrix} 0.8 & -0.6 & -0.6\\
-0.6 & 1.5 & -0.5 \\
-0.6 & -0.5 & 1.2\end{bmatrix}, 
\bm y=\begin{bmatrix} -0.8\\
-0.3 \\
-0.7\end{bmatrix}.
\end{equation*}
In Table~\ref{tab1}, we show values of $f_{ML}$, $f_{G}$, and $f_{LM}$, where $f_G$ represents the objective function for G-SB~(\ref{eq_g}).
We set $\lambda_g = 1$ for $f_G$ and $\lambda=1$ for $f_{LM}$.
The global minimum of these functions is $\bm x = [1,1,1]^T$. In addition, $f_{ML}$ and $f_{G}$ has a local minimum $\bm x = [-1,-1,-1]^T$ whereas $f_{LM}$ has no local minima.
This implies that SB minimizing $f_{ML}$ and $f_G$ possibly converges to the local minimum, resulting in detection error, but the SB with $\nabla f_{LM}$ is not the case.

\begin{figure}
\centerline{\includegraphics[width=\columnwidth]{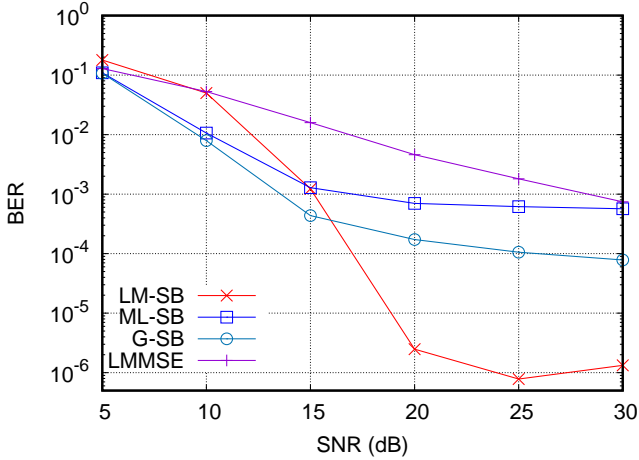}}
\caption{BER as a function of SNR when $N_t=N_r=16$.}\label{fig_1}
\end{figure}

We verify the superiority of the proposed SB in a massive MIMO sytem.
Figure~\ref{fig_1} shows the bit error ratio (BER) of variants of SB; 
ML-SB minimizing $f_{ML}$, G-SB minimizing $f_{G}$, and proposed LM-based SB (LM-SB) minimizing $f_{LM}$.
The antenna size is $N_t=N_r=16$. 
LMMSE represents the detection performance of the LMMSE detector. 
Following~\cite{SB2,SB3}, we set $\Delta=1$, 
$c_0=2[(N-1)/\sum_{i,j}J_{ij}^2]^{1/2}$ in~(\ref{eq_SB1}) and (\ref{eq_SB2}), 
$\lambda_g = 0.5$ for G-SB, and $\lambda = 1$ for LM-SB.
The number of iterations is fixed to $T=50$, and we set $a(t)= t/(\Delta T)$.
We find that all SB detectors are superior to the LMMSE detector. 
The ML-SB detector performs excellently in the low SNR region but shows an error floor in the high SNR region because of local minima. 
The G-SB detector also shows an error floor, although its detection performance is better than the ML-SB detector.
The proposed LM-SB detector shows performance improvement due to the local minima elimination
in the high SNR region.
The drawback of the LM-SB detector is the difficulty of choosing the parameter $\lambda$ and performance degradation in the low SNR region.

\section{Deep Unfolded Simulated Bifurcation}\label{sec_DUSB}
In the last section, we numerically show that the LM-SB detector using an LMMSE-like matrix improves MIMO detection performance. 
However, the choice of parameters such as $\lambda$, $\Delta$, and $c_0$ remains a critical issue and  directly affects the detection performance. 
This section aims  to construct DU-based LM-SB to improve detection performance with fewer iterations by tuning those parameters using training data.

\subsection{Differentiable SB and DU}
\begin{figure}
\centerline{\includegraphics[width=\columnwidth]{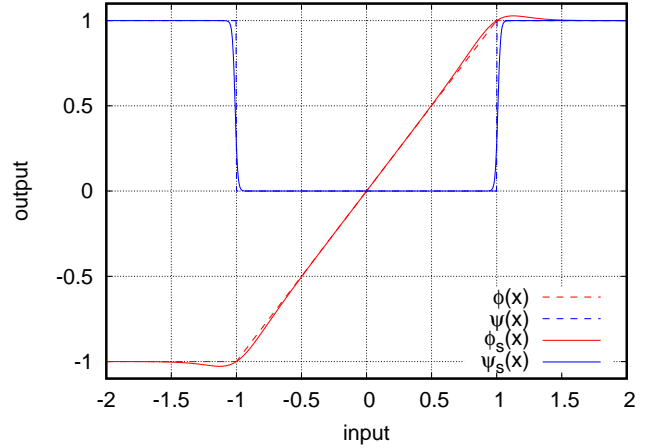}}
\caption{The plot of the clipping function $\phi(x)$ and ``square-well'' function $\psi(x)$, and their differentiable counterparts $\phi_s(x;\Lambda)$ and $\psi_s(x;A,B)$ with $\Lambda=10$, $A=100$ and $B=1.01$.}\label{fig_2}
\end{figure}

Here, we consider applying the notion of DU to the SB detector to tune internal parameters. 
To use the back-propagation, we need to make the update rule differentiable. 
The modified update rule is given by
\begin{align}
&\bm{\tilde x}(k) = \bm{x}(k) + \Delta_k \bm y(k), \label{eq_dSB1}\\
&\bm{\tilde y}(k) = \bm{y}(k)- \Delta_k \left[(1-a(k)) \bm{\tilde x}(k) 
+ \eta c_0 \left(\bm J\bm{\tilde x}(k)+\bm h\right)\right], \label{eq_dSB2}\\
&\bm{x}(k+1) = \phi_s(\bm{\tilde x}(k);\Lambda), \label{eq_dSB3}\\
&\bm{y}(k+1) = \bm{\tilde y}(k)[1 -  \psi_s(\bm{\tilde x}(k);A,B)], \label{eq_dSB4}
\end{align}   
where $a(k) = \sum_{l=0}^{k} \Delta_l/\sum_{l=0}^{T-1} \Delta_l$ ($k=0,\dots,T-1$) and 
$c_0=2[(N-1)/\sum_{i,j}J_{ij}^2]^{1/2}$.
We introduce step-size parameters $\{\Delta_k\}_{k=0}^{T-1}$ depending on the iteration index $k$, and 
scalar parameter $\eta$ controlling the strength of the term including $\bm J$ and $\bm h$ in (\ref{eq_dSB2}).
The functions $\phi_s(x:\Lambda)$ and $\psi_s(s;A,B)$ are defined as 
\begin{align}
 \phi_s(x;\Lambda)&:= \frac{1}{\Lambda}\left(\mathrm{sw}(\Lambda(x +1 )) 
\!-\! \mathrm{sw}(\Lambda(x-1 ))\right) \!-\!1, \label{eq_dSB3_1}\\
\!\psi_s(x ;A,B) &:= \sigma( A (|x| -B) ), \label{eq_dSB4_1}
\end{align}   
where $\mathrm{sw}(x):=x\sigma(x)$ is the Swish function and $\sigma(x):=1/(1+e^{-x})$ is the sigmoid function. 
They are differentiable and continuous functions corresponding to the clipping function $\psi(x)=-1$ ($x<-1$), $x$ ($|x|\le 1$),  $1$ ($x>1$) and ``square-well'' function $\psi(x):= 1$ ($|x|\ge 1$), $0$ ($|x|\le 1$) as shown in Fig.~\ref{fig_2}, respectively.  
These functions are applied to a vector element-wisely in (\ref{eq_dSB3}) and  (\ref{eq_dSB4})  to make the condition part below (\ref{eq_SB2}) differentiable.

The proposed deep unfolded SB named DU-LM-SB uses (\ref{eq_JH}) as $\bm J$ and $\bm h$.
It has trainable parameters $\{\Delta_k\}_{k=0}^{T-1}$, $\eta$, and $\lambda$ in $\bm U_\lambda$. 
The number of trainable parameters is $T+2$, which is constant to the antenna sizes $N_t$ and $N_r$.

\subsection{Simulation Setting}

We describe the simulation settings. 
In the MIMO system (\ref{eq_rec}), the SNR is given by $\mathrm{SNR}=10\log_{10}\frac{n}{\sigma_w^2}$. 
All detected signals are thresholded to $\mathbb{S}=\{1,-1\}$ for calculating the BER. 

The DU-LM-SB is implemented by PyTorch 2.0~\cite{PyTorch}. 
It is trained in a supervised manner using randomly generated pairs of transmitted and received signals $\{(\bm x,\bm y)\}$. 
Trainable parameters are updated by the Adam optimizer~\cite{Adam} with a learning rate of $2\times 10^{-4}$ to minimize the MSE loss function.
In each parameter update, 1000 mini-batches of size 2000 are fed. 
A channel matrix $\bm H$ is generated for each mini-batch.
We set $T=10$, and the initial values of the trainable parameters were $\Delta_k=1.0$ and $\eta=\lambda=1$. The parameters for $\phi_s$ and $\psi_s$ are set to $\Gamma=10,A=100$, and $B=1.01$.

We also examined the MMSE detector and TPG detector~\cite{TPG2} as baseline algorithms.
The TPG detector is an MIMO detector based on DU and projected gradient descent. 
The detector with $T$ iterations has $2T+1$ trainable parameters. 
It was trained and executed under the same conditions as in~\cite{TPG2}.
The number of iterations of the TPG detectors was set to $T=10$, as in the case of the DU-LM-SB detector.
The time complexity of the DU-LM-SB and TPG detectors is $O(N_t^3)$.

\begin{figure}
\centerline{\includegraphics[width=\columnwidth]{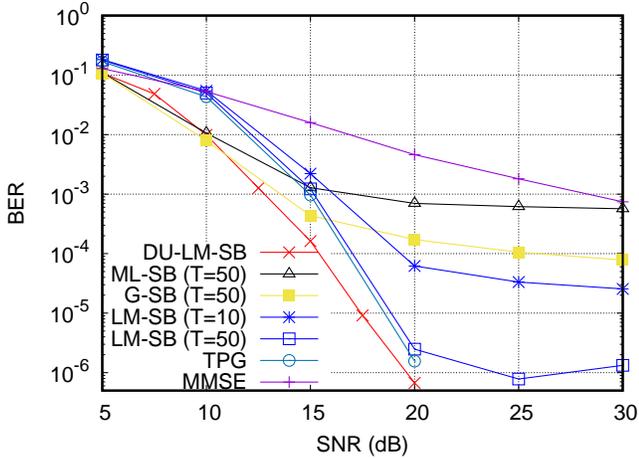}}
\caption{BER performance of MIMO detectors when $N_t=N_r=16$. Note that the estimated BERs of DU-LM-SB and TPG  are zero when the SNR is larger than $20$dB. }\label{fig_n2}
\end{figure}

\subsection{Numerical Results}
Here we demonstrate the detection performance of the proposed algorithm. 

Figure~\ref{fig_n2} shows the detection performance of DU-LM-SB and baseline detectors when $N_t=N_r=16$; ML-SB with $T=50$, G-SB with $T=50$ and $\lambda_g=0.5$, LM-SB with $T=10,50$ and $\lambda=1$, TPG, and MMSE detectors. 
We find that the BER of the LM-SB detector with $T=10$ is larger than that with $T=50$ but still smaller than the MMSE detectors. 
The trained TPG detector shows the detection performance close to the LM-SB detector with $T=50$ but  has no error floor in the high SNR region. 
The ML-SB and G-SB detectors perform excellently in the low SNR region, but their performance degrades when the SNR is above $20$dB.  
The proposed DU-LM-SB detector performs best and has no error floor by tuning its trainable parameters. 
The gain of the DU-LM-SB detector against the LM-SB ($T=50$) and TPG detectors are about $2.5$dB when BER$=10^{-3}$.

\begin{figure}
\centerline{\includegraphics[width=\columnwidth]{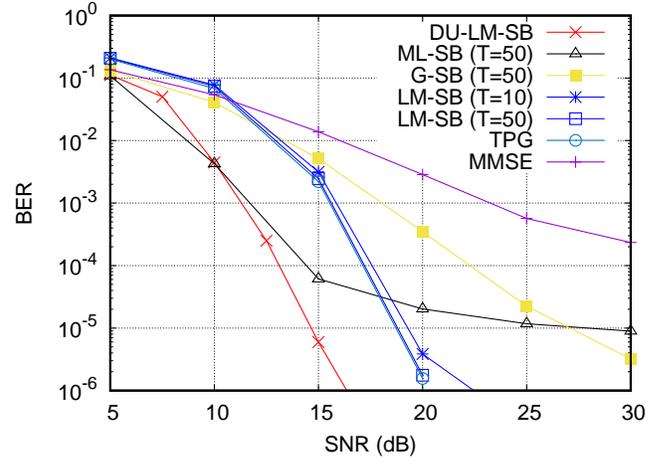}}
\caption{BER performance of MIMO detectors when $N_t=N_r=32$. }\label{fig_n3}
\end{figure}

We also show the detection performance of DU-LM-SB and baseline detectors when $N_t=N_r=32$ in Figure~\ref{fig_n3}. 
In this case, the performance of the LM-SB detector with $T=10$ is close to that of LM-SB with $T=50$ and TPG detectors, although it still has an error floor. 
The proposed DU-LM-SB detector shows improved performance 
compared with other detectors in both low and high SNR regions. 
The gain of the DU-LM-SB detector against the LM-SB ($T=50$) and TPG detectors is about $5$dB when BER$=10^{-5}$.
    
\section{Concluding Remark}\label{sec_con}

In this letter, we proposed two SB-based detectors for MIMO systems. 
SB is a quantum-inspired algorithm for solving QUBOs, including MIMO signal detection.  
We first introduced the LM-SB detector that possibly eliminates local minima and  improves the error floor observed in existing SB detectors. 
In addition, by combining it with DU, we proposed a trainable quantum-inspired MIMO detector called DU-ML-SB that has a constant number of training parameters controlling the dynamics of SB. 
The numerical simulations show that the proposed detectors perform better than existing SB-based detectors. 
Additionally, the DU-ML-SB detector is superior to another conventional trainable MIMO detector, which shows the potential of quantum-inspired algorithms and DU. 

The results suggest that a simple modification by the LM algorithm can improve the performance of other QUBO solvers. 
It is also suggested that DU is effective in not only iterative optimization algorithms but also solvers defined as dynamical systems. 
It is an interesting topic to construct dynamical systems for signal processing~\cite{NW1,NW2} and combine them with DU.  
Investigation of the performance of the proposed detectors for massive overloaded MIMO systems and MIMO systems with a higher-order modulation is another future task.

\end{document}